%% file: main.tex
\documentclass[nonacm, sigconf]{acmart} 

\input{aliases} 
\input{ethos.tex}

\usepackage{xspace}
\usepackage{multirow}
\usepackage{float}
\usepackage{listings}
\usepackage{natbib}

\definecolor{jsonpurple}{rgb}{0.5,0,0.35}
\definecolor{jsonblue}{rgb}{0.0,0.0,0.6}

\begin{document}

\title[Scalable Characterization of Message-Based Job Scams]{\sys: Scalable Characterization of Message-Based Job Scams}

\author{Abisheka Pitumpe}
\affiliation{
  \institution{Stony Brook University}
  \state{New York}
  \country{USA}}
\email{abisheka.pitumpe@cs.stonybrook.edu}

\author{Amir Rahmati}
\affiliation{
  \institution{Stony Brook University}
  \state{New York}
  \country{USA}}
\email{amir@rahmati.com}

\begin{abstract}
Job-based smishing scams, where victims are recruited under the guise of remote job opportunities, represent a rapidly growing and understudied threat within the broader landscape of online fraud. In this paper, we present \sys, the first scalable, end-to-end measurement pipeline designed to systematically engage with, analyze, and characterize job scams in the wild. \sys combines large language models (LLMs), automated browser agents, and infrastructure fingerprinting tools to collect over 29,000 scam messages, interact with more than 1900 scammers, and extract behavioral, financial, and infrastructural signals at scale. We detail the operational workflows of scammers, uncover extensive reuse of message templates, domains, and cryptocurrency wallets, and identify the social engineering tactics used to defraud victims. Our analysis reveals millions of dollars in cryptocurrency losses, highlighting the use of deceptive techniques such as domain fronting and impersonation of well-known brands. \sys demonstrates the feasibility and value of automating the engagement with scammers and the analysis of infrastructure, offering a new methodological foundation for studying large-scale fraud ecosystems.
\end{abstract}




\maketitle

\input{sections/introduction}

\input{sections/background}

\input{sections/anatomy}
\input{sections/system-design}

\input{sections/analysis}

\input{sections/financial-loss}

\input{sections/discussion}
\input{sections/related-work}

\input{sections/conclusion}


\bibliographystyle{ACM-Reference-Format}
\bibliography{ref}

\appendix
\section{Ethical Considerations}
\input{sections/ethics}


\input{sections/appendix}

\end{document}

%% file: aliases.tex
\newcommand{\measurementPeriod}{10 month}

\newcommand{\biggestInitialMessageCluster}{600}
\newcommand{\totalMessages}{29,209}
\newcommand{\totalScammersNumbers}{7028}
\newcommand{\totalSMS}{6559}
\newcommand{\totalWhatsApp}{139}
\newcommand{\totalTelegram}{330}
\newcommand{\totalSuccessfulInteractions}{521}
\newcommand{\totalFailures}{1944}

\newcommand{\biggestclusterscreenshot}{42}

\newcommand{\sys}[0]{Anansi\xspace}

\newcommand{\totalBTCwallets}{262}
\newcommand{\totalETHwallets}{209}
\newcommand{\uniqueBTCwallets}{205}
\newcommand{\uniqueETHwallets}{149}
\newcommand{\totalUniqueWallets}{354}

\newcommand{\totalBitcoinUSD}{\$514,528}
\newcommand{\totalEthUSD}{\$11,768,730}
\newcommand{\totalLossUSD}{\$12,283,258}

\newcommand{\totalwebsites}{218}
\newcommand{\uniqueScamWebsites}{135}
\newcommand{\uniqueIPs}{80}
\newcommand{\cloudflarePercent}{44.83\%}
\newcommand{\integenPercent}{5.88\%}

\newcommand{\target}{454}
\newcommand{\costco}{385}
\newcommand{\amazon}{417}
\newcommand{\structube}{195}
\newcommand{\qualtrics}{31}
\newcommand{\sony}{18}
\newcommand{\warnerbros}{56}
\newcommand{\linkedin}{357}
\newcommand{\ssense}{62}
\newcommand{\dsl}{86}

\newcommand{\judy}{264}
\newcommand{\maya}{234}
\newcommand{\linda}{130}
\newcommand{\elowen}{123}
\newcommand{\joshblair}{60}
\newcommand{\jasmine}{312}
\newcommand{\darlene}{69}
\newcommand{\monica}{38}
\newcommand{\isabella}{60}
\newcommand{\elizabeth}{19}

\newcommand{\totalTaskScamsPercent}{67.51\%}
\newcommand{\totalYouTubeScamsPercent}{18.29\%}
\newcommand{\totalAppStoreScamsPercent}{14.20\%}

\newcommand{\bitcoinMedianTxns}{11}
\newcommand{\bitcoinAvgTxns}{12.9}
\newcommand{\bitcoinMinTxns}{1}
\newcommand{\bitcoinMaxTxns}{24}

\newcommand{\bitcoinMedianIncome}{\$3,240}
\newcommand{\bitcoinAvgIncome}{\$5,474}
\newcommand{\bitcoinMinIncome}{\$30}
\newcommand{\bitcoinMaxIncome}{\$69,106}

\newcommand{\ethMedianTxns}{34}
\newcommand{\ethAvgTxns}{266}
\newcommand{\ethMinTxns}{1}
\newcommand{\ethMaxTxns}{5,090}

\newcommand{\ethMedianIncome}{\$10,105}
\newcommand{\ethAvgIncome}{\$79,518}
\newcommand{\ethMinIncome}{\$1}
\newcommand{\ethMaxIncome}{\$2 million}

%% file: ethos.tex
\usepackage{color}
\usepackage{xcolor}
\usepackage{xspace}
\usepackage{url}
\usepackage[font={footnotesize},labelfont=bf]{caption,subfig} 
\usepackage{pifont} 
\usepackage{amsmath}
\usepackage{todonotes} 

\usepackage{enumitem} 
\setlist{nosep} 
\setlist[itemize]{leftmargin=*} 

\usepackage{wrapfig}
\usepackage{rotating} 

\usepackage{booktabs} 

\usepackage{array} 
\newcommand{\PreserveBackslash}[1]{\let\temp=\\#1\let\\=\temp}
\newcolumntype{C}[1]{>{\PreserveBackslash\centering}p{#1}}
\newcolumntype{R}[1]{>{\PreserveBackslash\raggedleft}p{#1}}
\newcolumntype{L}[1]{>{\PreserveBackslash\raggedright}p{#1}}

\newcommand{\paratitle}[1]{\noindent\textbf{\textit{#1.}}\xspace}

\newcommand{\etal}[0]{\emph{et~al.}\xspace}
\newcommand{\eg}[0]{\emph{e.g.,}\xspace}

\usepackage{cleveref} 


%% file: sections/introduction.tex
\section{Introduction}~\label{sec:introduction}

Online fraud has recently reached unprecedented levels, with devastating financial and psychological consequences~\cite{Norris2019PsychologyInternetFraud,psychological_romance,10.1145/3449115, Beech_2026} for victims worldwide. The Federal Bureau of Investigation (FBI) has revealed that Americans lost approximately \$2.8 billion to cryptocurrency fraud in 2024, marking a 71\% increase from the previous year~\cite{FBI_IC3_2024}. Among the most sophisticated and damaging forms of online fraud are ``pig butchering" scams, a translation of the Chinese phrase \textit{sha~zhu~pan}, which refers to an online fraud tactic in which scammers gradually gain the trust of victims before defrauding them, similar to fattening a pig before slaughter~\cite{pigbutchering-history}. 

Despite the massive scale and growing prevalence of such fraud, systematic measurement and analysis of these operations remain limited, with most efforts focused on web-based scams. Notable works in this area include Vasek~\etal~\cite{vasek}, who conducted early work characterizing Bitcoin-based scams, identifying common patterns in Ponzi schemes and fraudulent investment platforms; Li~\etal\cite{double_or_nothing}, who developed CryptoScamTracker for identifying and tracking cryptocurrency giveaway scams that exploit social media platforms; Liu~\etal\cite{Liu2024_GiveAndTake} who conducted an end-to-end investigation of giveaway scam conversion rates; and Muzammil~\etal~\cite{muzammil2025babylon}  who introduced Crimson for detecting cryptocurrency investment scam websites. 

Recent work has also examined message-based scams. Nahapetyan \etal ~\cite{nahapetyan2024sms} analyzed 68K phishing messages and the SMS phishing infrastructure, uncovering systematic abuse in domain registration and hosting ecosystems and identifying key evasion techniques used by scammers. Agarwal~\etal~\cite{agarwal2024sms} propose a taxonomy of SMS scams based on analysis of millions of messages, identifying eight distinct scam categories and documenting a 27-fold increase in SMS phishing since 2020. 
However, significant gaps remain in our understanding of modern message-based fraud ecosystems. First, existing work is retrospective and does not provide a real-time, end-to-end pipeline for capturing, analyzing, and engaging with scam messages. Second, existing approaches lack message-level clustering techniques that can track campaign evolution across reused linguistic templates, contact details, and cryptocurrency wallet addresses. Finally, previous work has focused solely on textual content and has not integrated multimodal analysis, combining text, images, and scammers' behavioral patterns within a unified framework.

To address these limitations, we present \sys{}\footnote{\sys is a popular trickster figure in West African and Caribbean folklore, often depicted as a spider.}, the first scalable, end-to-end measurement pipeline designed to systematically engage with, analyze, and characterize job scams in the wild. 
\sys creates a semi-automatic pipeline that uses web crawlers and crowdsourcing to continuously monitor and flag potential scammers. Next, \sys uses custom LLM agents to engage with these scammers, luring them into disclosing information such as message templates, websites, and wallet addresses. This approach provides deeper insights into the tactics, infrastructure, and reuse patterns characteristic of modern pig butchering campaigns, as well as a glimpse into the extent of financial loss caused.

During \sys{}'s \measurementPeriod{} measurement period, \sys collected and analyzed \totalMessages{} messages from \totalScammersNumbers{} scammer phone numbers across SMS, WhatsApp, and Telegram platforms. We extracted transaction data from \totalUniqueWallets{} scammer-owned Ethereum and Bitcoin wallet addresses representing an estimated financial loss of \totalLossUSD{}. Furthermore, we discover commonalities among scammers and cluster properties to show the interconnectedness of scam syndicates.

Our contributions are summarized as follows:
\begin{itemize}
    \item We developed \sys, the first scalable, semi-automated pipeline to engage with and analyze job-based smishing scams in real time.
    \item Using \sys, we collected and analyzed 29,000+ messages and 1,900+ scammer interactions, revealing patterns in scammer behavior, messaging templates, and infrastructure reuse.
    \item By analyzing websites and wallet addresses shared by the scammers, we uncovered \$12M+ in crypto losses, linked scams to specific wallet addresses, domains, and hosting providers. We further exposed deceptive techniques, including domain fronting and wallet rotation, and clustered campaigns by infrastructure and behavior, demonstrating coordinated scam operations across multiple platforms.
\end{itemize}

%% file: sections/background.tex
\section{Background}

\paratitle{Pig-Butchering Scams}
 According to the FBI's Operation Level Up~\cite{fbi_operation_level_up_2025}, 

 pig butchering schemes resulted in over \$5.8 billion in losses in 2024 alone.
Offenders initiate contact on social media platforms (such as Facebook and Instagram), professional networking sites (\eg LinkedIn), dating services (\eg Hinge, Tinder), and increasingly on communication platforms such as WhatsApp, Telegram, and WeChat ~\cite{FinCEN_Alert}. Many victims are also targeted via unsolicited SMS messages or phone calls, and some are even added to group chats designed to promote fraudulent investment schemes ~\cite{investorAlert}.

Once a victim engages, the scam unfolds in a calculated sequence of social engineering steps~\cite{coinbase_psa}. Initially, the target is approached with the promise of easy income or a high-yield investment opportunity. If they express interest, they are guided through what appears to be a formal interview or onboarding process, often involving the submission of personal data or account creation. The scammer gradually builds trust, culminating in a financial request framed as a refundable fee for job training, software, or work equipment. By the time payment is made or credentials are shared, the scammer typically disappears, leaving the victim defrauded. As of July 2025, 64 victims have been referred to an FBI victim specialist for suicide intervention, which shows how devastating the effects of these scams can be ~\cite{fbi_operation_level_up_2025}.

\paratitle{Job Scams}
Job scams, a subset of pig-butchering scams, represent an emerging and increasingly prevalent form of online fraud~\cite{FTC_TaskScams2024,MOAA_TaskScams2025}. 

In job scams, perpetrators offer what appear to be legitimate job opportunities, promising generous compensation for completing simple online tasks. These tasks often include rating products on fraudulent websites designed to resemble reputable retailers, liking videos on platforms like YouTube, or engaging with other forms of social media content. Scammers typically contact victims via SMS, WhatsApp, or Telegram, posing as recruiters from well-known companies to gain their trust.

What makes job scams especially deceptive is their exploitation of real economic hardship. They often target individuals who are unemployed, underemployed, or in need of supplemental income, making the promise of easy earnings particularly appealing. The financial impact of these schemes has grown dramatically, with job scams costing a reported \$41 million in cryptocurrency losses alone in the first six months of 2024, up from \$21 million in all of 2023, according to the FTC~\cite{MOAA_TaskScams2025}.

%% file: sections/anatomy.tex

\section{The Anatomy of a Task}~\label{sec:anatomy}
\begin{figure}[t]
    \centering    \includegraphics[width=\columnwidth]{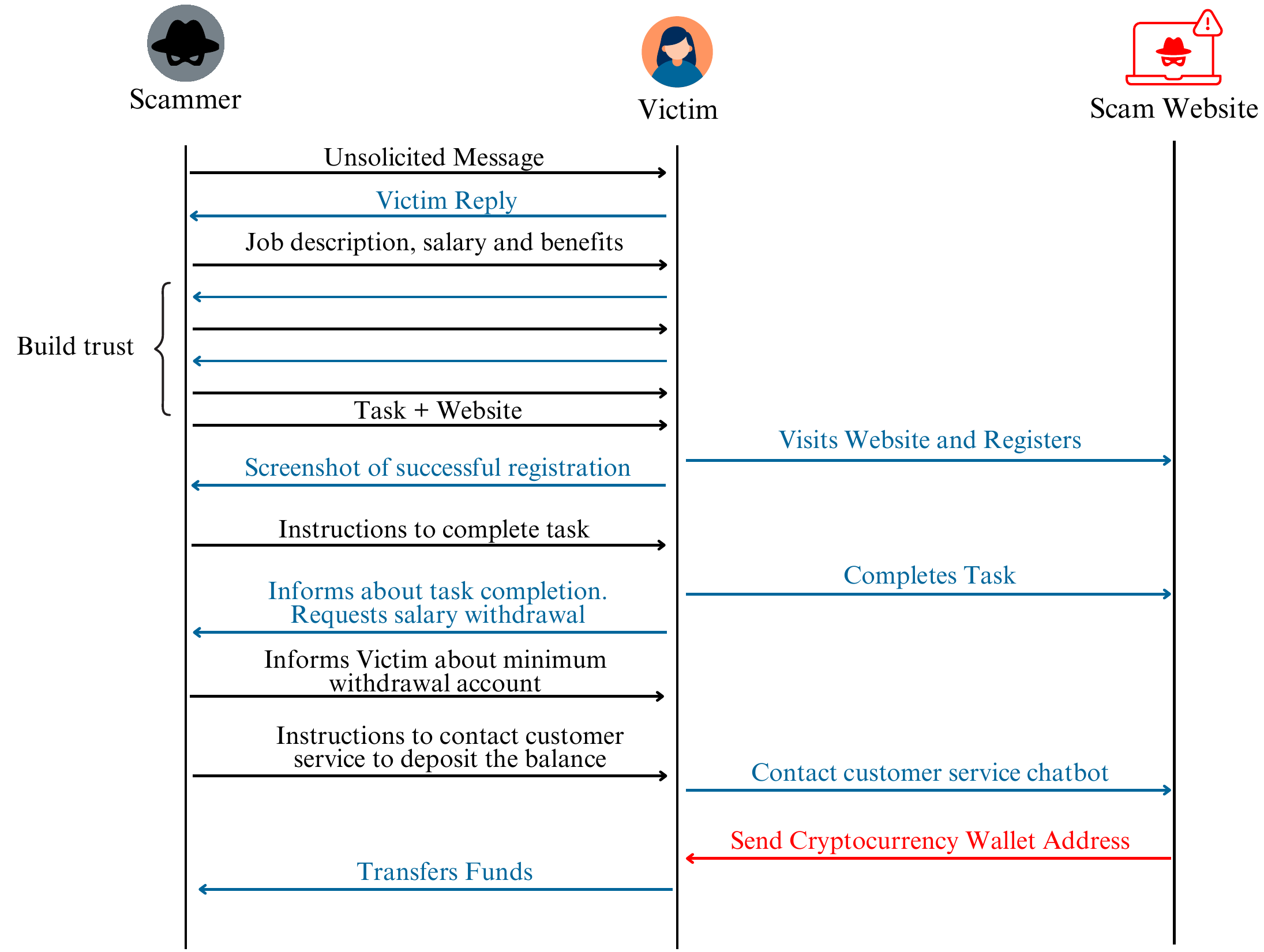}
    \caption{Anatomy of a Task Scam }\label{fig:anatomy}
\end{figure}

\begin{figure}[t]
    \centering
    \includegraphics[width=0.5\textwidth]{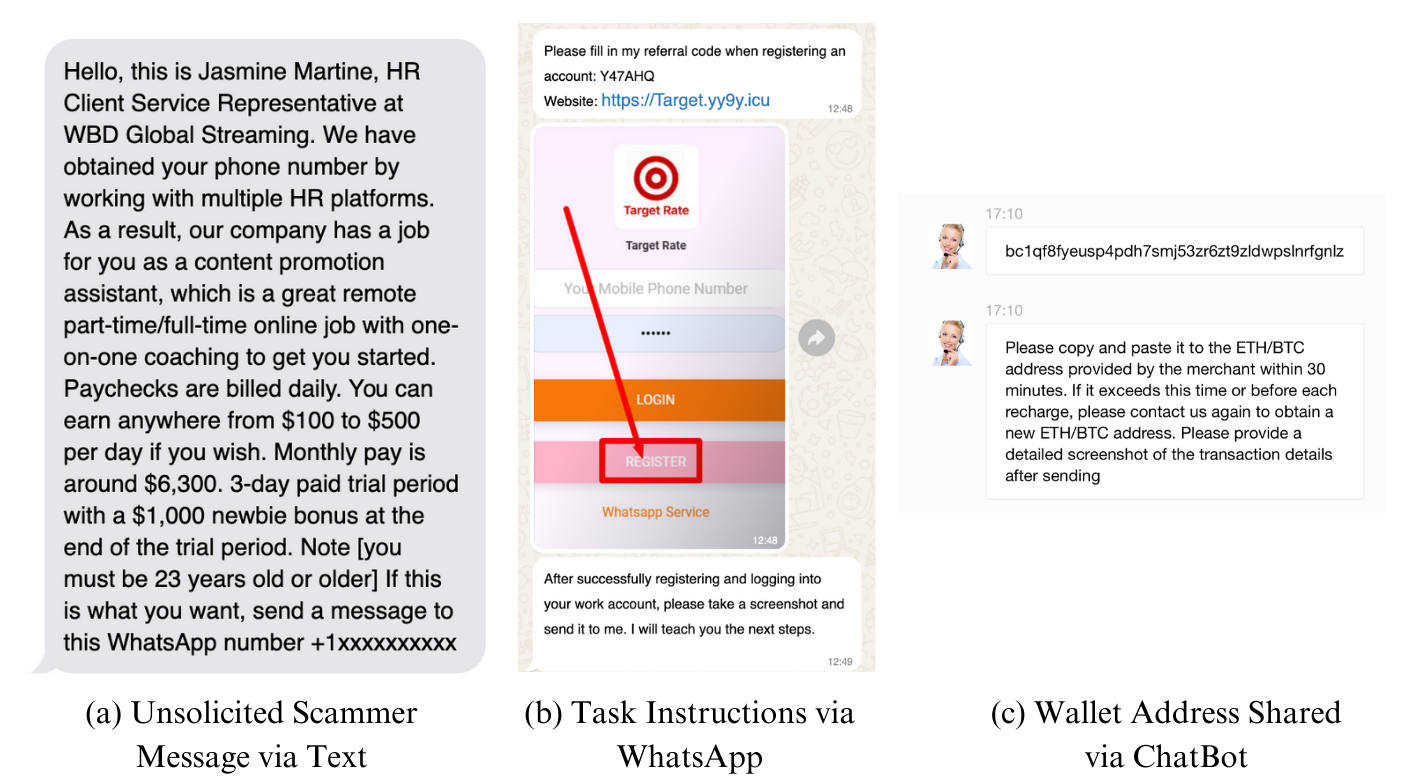}
    \vspace{2ex}
    \caption{Examples of Scammer Messages}\label{fig:types_of_scam_messages}
\end{figure} 
Job-based scams follow a systematic, multi-stage progression designed to gradually build victim trust while maximizing financial extraction~\cite{doi:10.1177/26338076241248176}. Figure~\ref{fig:anatomy} illustrates the complete workflow from initial contact through final payment extraction.

\paratitle{Initial Contact}
Scammers send unsolicited SMS, WhatsApp, or Telegram messages claiming to have found the victim's resume on legitimate platforms like Indeed or LinkedIn. Messages promise \$250-\$500 daily for minimal remote work.

\paratitle{Trust Building}
Scammers adopt HR personas, inquire about victims' backgrounds, and may provide fabricated business licenses and salary structures to appear legitimate.

\paratitle{Platform Handoff and Training}~\label{subsec:platform_handoff}
Victims are referred to a ``trainer'' on WhatsApp or Telegram with a unique referral code. Trainers claim sessions are recorded and require 30-60 minutes of the victim's time.

\paratitle{Account Registration and Task Completion}
Victims register on scam websites mimicking legitimate businesses (Figure~\ref{fig:types_of_scam_messages}(b)), providing personal information and the referral code. After completing 20-40 review tasks, victims see ``earnings'' displayed on dashboards.

\paratitle{Payment Extraction}
When victims attempt to withdraw funds, scammers impose minimum-balance requirements, forcing them to deposit cryptocurrency first. Scammers initially return small deposits with ``profits'' to build trust, then gradually request larger amounts (\$100, \$300, \$800+) over weeks or months. Each successful return reinforces legitimacy.

\paratitle{Final Payment and Ghosting}
After deposits escalate to a large sum, scammers request a final large payment, then disappear—claiming technical issues or simply blocking the victim. Victims lose access to all accumulated funds.

%% file: sections/system-design.tex
\section{\sys Design}~\label{sec:system-design}

\begin{figure*}[h!]
    \centering
    \includegraphics[width=1\textwidth]
    {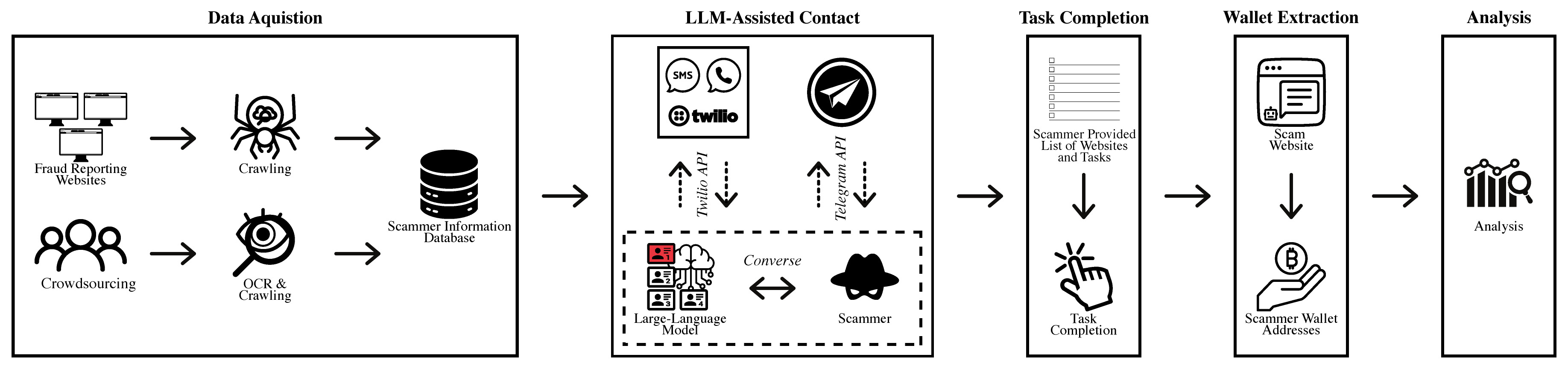}
    \vspace{2ex}
    \caption{A schematic representation of \sys{}'s pipeline}
    \label{fig:sys-diagram}
\end{figure*}

Figure~\ref{fig:sys-diagram} provides a high-level overview of \sys{}. Our system comprises five core modules operating within its pipeline to collect, process, and analyze scam-related data at scale. These modules are: (i)~Data Acquisition, (ii)~LLM-Assisted Contact Initiation, (iii)~Task Completion, (iv)~Wallet Extraction, and (v)~Analysis. We describe each of these components in detail below.

\subsection{Data Acquisition}

Acquiring comprehensive scam data presents significant challenges due to the distributed and ephemeral nature of scam operations. Scammers frequently change phone numbers, web domains, wallet addresses, and employ various techniques to avoid detection. Additionally, scam data is scattered across multiple heterogeneous sources, each with different formats, access patterns, and data quality levels. Furthermore, a critical temporal gap exists between when scammers initially contact victims, when victims recognize and report these interactions, and when these reports are processed. While this gap cannot be eliminated, \sys{}'s automated data acquisition pipeline reduces processing time and enables faster response to newly reported scams.

The Data Acquisition module systematically collects scammer phone numbers and message content from two sources: (1) public reporting portals and crowdsourced databases, and (2)~direct submissions from our research team and collaborators.

For web-based data, we use Selenium-based automated scrapers to extract relevant information and store it in a structured CSV format. These datasets are then ingested into a custom-built dashboard for inspection and further analysis. 

One primary data source is the Better Business Bureau’s scam reporting portal~\cite{better_business}, where members of the public report details about scammers. Another key source is Smishtank~\cite{timko2024smishing}, a crowdsourced repository of smishing messages. We selected these as our main sources as they are the only publicly available datasets with continuous active user submissions on employment scams. 
From these sources, we extract both message content and sender information, focusing on messages submitted under the ``Employment" scam type~\cite{BBB_ScamTips}.

In addition to textual data scraped from websites, we process scam messages received as screenshots from our research team and collaborators. We recruited participants through department mailing lists and our collaboration network, and asked them to forward suspicious messages they encountered in their daily communications. These submissions come from various messaging platforms, including SMS, WhatsApp, and Telegram. The images are processed using Optical Character Recognition (OCR) to extract phone numbers and message content. The extracted data are similarly converted to CSV format and imported into the dashboard, ensuring unified data representation across all sources. This direct submission approach provides access to fresh scam attempts that may not yet be reported in public portals, helping to reduce the temporal gap between scammer activity and our detection capabilities.

\subsection{LLM-Driven Scammer Engagement}
Once the scammer's contact information has been obtained, \sys initiates contact via dedicated research phone numbers provided by Twilio's Programmable Messaging Service. 
\sys{} automatically initiates contact and uses a dictionary of pre-determined messages to reach out to scammers. This module is designed to simulate real user engagement with scammers across different platforms (\eg SMS, Telegram, WhatsApp), while maintaining plausible yet controlled interactions that prompt scammers to reveal additional information. \sys{} leverages custom profiles, fine-tuned prompts, and dynamic templates to generate replies tailored to the content and tone of scammer messages.

Successful scammer engagement requires systematic creation and maintenance of believable victim personas. Our system implements a dedicated profile generation module. For each new scammer engagement session, \sys{} generates a comprehensive victim profile that includes detailed demographic and financial information, often asked by scammers. Once a profile is assigned to a particular scammer conversation, the system maintains a persistent profile database along with other metadata about the conversation. This ensures that identical persona characteristics are continued throughout the entirety of the conversation, even if the conversation migrates across different platforms or gets passed between multiple scammers (\eg Adina from the Costco Recruiting team). While \sys leverages custom profiles, fine-tuned prompts, and dynamic templates to generate replies based on the content and tone of the scammer messages, recognizing that overly mechanical responses could raise suspicion among scammers. Therefore, we continuously monitor all LLM-generated conversations. If the LLM responds in ways that could appear suspicious or artificial, human intervention is triggered to ensure the conversation maintains authenticity and continues to elicit information effectively from scammers.

\sys is designed to engage scammers across different communication channels, including SMS, WhatsApp, and Telegram. Message delivery and reception for SMS and WhatsApp are handled through Twilio~\cite{twilio}, a programmable SMS API, with hooks into the pipeline to update message status and log responses. Telegram interactions are managed through the platform's bot API~\cite{telegram_api}.
This enables large-scale, semi-automated conversations with scammers while preserving the flexibility to escalate scammer conversations for human intervention if necessary.

\subsection{Task Completion}

Upon successful engagement, scammers typically instruct the victim to perform the required ``tasks'' to generate the presumed earnings. We will discuss the different types of scams our system encountered in Section \ref{sec:types_of_scams}. In the majority of the scams we encountered, scammers provide specific websites where alleged ``tasks" are to be performed, along with referral codes or invite links that enable account creation. \sys systematically visits these provided websites using automated browsing tools Selenium~\cite{selenium} and ChromeDriver~\cite{chromedriver}, registers accounts using the supplied credentials, and logs in to emulate a typical victim workflow. Throughout this process, we capture detailed screenshots of all accessible pages, with particular focus on account creation, task interfaces, and payment mechanisms.
Scammers maintain frequent contact with victims throughout the task, often requesting screenshots to verify account creation, task completion, and payment preparation.

\subsection{Wallet Extraction}~\label{subsec:wallet_extraction}
The Wallet Extraction module serves as the endpoint of our scammer engagement process, designed to elicit the complete sequence of scam instructions and ultimately obtain the cryptocurrency wallet addresses used for victim payments.

Acquiring cryptocurrency wallet addresses involves identifying and engaging with the customer support infrastructure that scam operations establish to facilitate victim payments. These support channels serve as the primary path for scammers to provide payment instructions and wallet addresses to victims.
Crucially, the scammer instructs victims to contact the website's customer support to obtain the appropriate cryptocurrency wallet address for payment. We systematically identify and document these customer support channels, which may be implemented as embedded web chat services, redirects to external messaging platforms such as WhatsApp or Telegram, or direct contact information for ``customer service" representatives.

We utilize OpenAI's Operator \cite{openai2025operator} and custom automation scripts built with Selenium \cite{selenium} to navigate these scam websites, extract customer support contact information, and initiate contact with support representatives. By explicitly requesting wallet addresses required for deposits or transaction completion, we obtain critical intelligence about the payment infrastructure used across different scam operations.
These wallet addresses serve as the foundation for subsequent financial analysis, enabling the tracking of funds flows across related scam campaigns.

%% file: sections/analysis.tex
\section{Analysis}~\label{sec:analysis}

\begin{figure*}
    \centering

    \includegraphics[width=.9\textwidth]{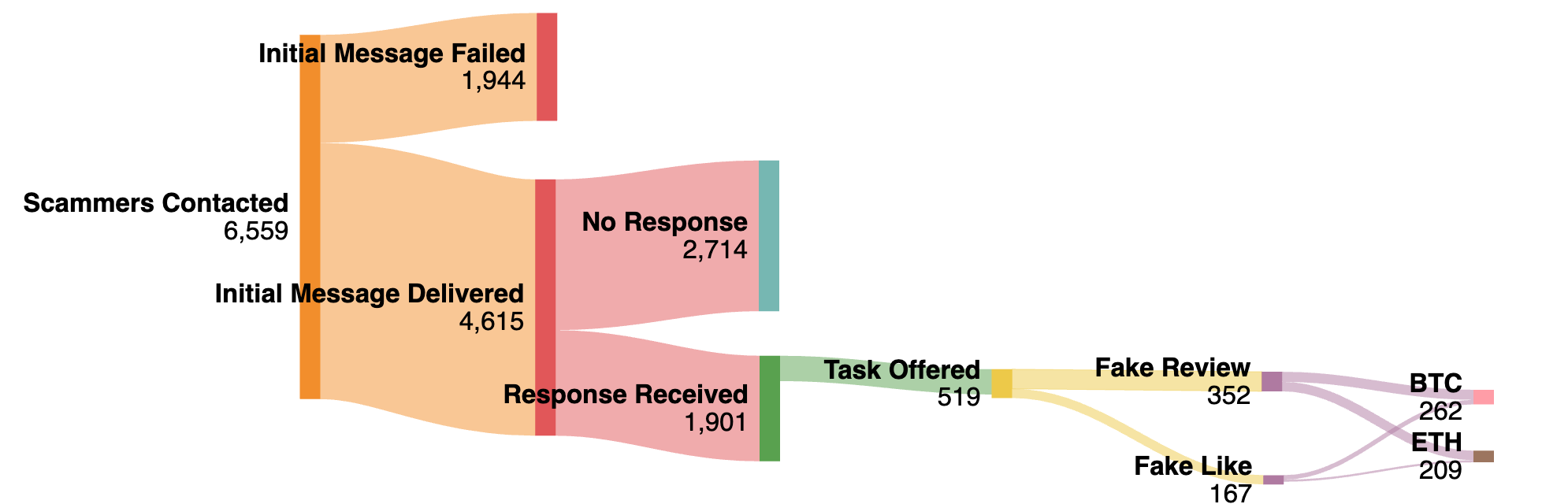} 
    \caption{Scammer Attrition Flow}
        \label{fig:sankey}
\end{figure*}

Over its 10-month measurement period (April 14th, 2025 - January 31st, 2026), \sys{} collected and analyzed \totalMessages{}  messages from a total of \totalScammersNumbers{} scammer phone numbers.  Of these numbers, 6,874 were discovered through public reporting portals, while 154 were obtained from direct submissions by the research team and collaborators. \sys{} initiated contact with these \totalScammersNumbers{} scammers across three platforms: \totalSMS{} via SMS, \totalWhatsApp{} via WhatsApp, and \totalTelegram{} via Telegram. Out of the contacted scammers, \sys{} successfully received responses and interacted with 1,901.
In this section, we first examine \sys performance in engaging scammers, completing tasks, and acquiring cryptowallet addresses. Next, we analyze collected data across multiple dimensions to understand modern job scam operations.

\subsection{Stage-wise Attrition Analysis}~\label{subsec:attrition}

Figure~\ref{fig:sankey} provides an overview of \sys attrition across various stages of its pipeline. 
\sys{} failed to initiate conversations with \totalFailures{} numbers due to several reasons. 693 phone numbers failed to deliver as the device was powered off~\cite{Twilio_Error30003}, while 692 numbers were landlines~\cite{Twilio_Error30005,Twilio_Error30006}. 141 failures were associated with WhatsApp restrictions~\cite{Twilio_Error63016,Twilio_Error63024,Twilio_Error63049}. A significant technical constraint emerged when trying to engage with scammers who provided WhatsApp contact information for their ``trainers" or supervisors. Meta's WhatsApp Business API, which we access via Twilio's integration, enforces strict policies that prevent business accounts from initiating conversations with users who have not previously messaged the business account. This policy is designed to improve the consumer experience, but it creates a barrier to our research methodology~\cite{twilio_whatsapp}.
To address this limitation, we developed a systematic workaround that leverages the existing communication channel with the initial scammer contact. When scammers provide WhatsApp numbers for trainers or support personnel, we craft carefully worded messages to the original point of contact, explaining that we are unable to initiate WhatsApp conversations due to account restrictions.
Specifically, we inform the initial scammer that our WhatsApp account has messaging limitations that prevent us from contacting new numbers directly. We request that they either: (1) have their trainer contact us first, or (2) provide alternative platforms like Telegram.

For scammer numbers where messages were successfully delivered but no response was received, we hypothesize that the numbers may have been deactivated or taken down during our study. A similar explanation may apply to scammers who initially responded but then ceased further engagement.

\subsection{ Message Characteristics}~\label{sec:message_characteristics}

We analyze patterns and commonalities in the initial outreach messages sent by scammers. Common indicators include urgent or overly enthusiastic language (\eg ``earn \$250 to \$500 a day''), unrealistic promises of high compensation for minimal effort, generic greetings, inconsistent branding, and directives to continue communication via alternative platforms such as WhatsApp. Many messages also contain poor grammar, typographical errors, and mismatches in sender identity in follow-up exchanges. 
These messages frequently reference well-known brands, as shown in Table~\ref{tab:brand-and-names}, claim to have found victims' resumes on recruitment sites. Our findings indicate that scammers employ mass-messaging strategies, distributing initial messages to large recipient lists to maximize response rates. This aligns with previous observations, such as those by DeLiema~\etal~\cite{deliema2024profiling}. \sys provides additional empirical evidence for these mass-messaging strategies, as when \sys{} contacts scammers (Section~\ref{sec:system-design}), it responds to messages originally sent to other users. Despite never having contacted our numbers, scammers consistently engage with our replies, suggesting they distribute messages broadly without tracking recipients and treat any response as a potential victim. 

Initial messages sent by scammers fall into three high-level categories: (1)~detailed job offers,  (2)~exploratory inquiries, and (3)~wrong number messages. The first category presents comprehensive job descriptions with specific salary ranges, work requirements, and company benefits. These messages often impersonate legitimate companies and promise high earnings (\eg \$250-\$500 daily) for minimal time commitments (\eg 60-90 minutes per day).

The second category consists of brief inquiries that gauge recipient interest without revealing job specifics. These messages introduce the sender as a recruiter who has supposedly reviewed the recipient's resume through online platforms, prompting recipients to express interest in learning more about high-paying remote opportunities.

Both message types reference well-known companies and claim to have accessed victims' resumes through established job platforms such as Indeed or LinkedIn to establish credibility during initial contact.

The third category consists of wrong-number texts~\cite{bitdefender_wrongnumber2025} where scammers send seemingly innocent texts addressed to someone else. These messages typically begin with casual personal content, such as ``Hey Sarah, are we still meeting for dinner tonight?" When recipients respond to clarify the mistake, scammers apologize for the error and often try to engage in a friendly conversation. Once engagement is established, scammers gradually transition the conversation toward job opportunities or investment advice, often claiming to work in lucrative fields or mentioning their financial success in passing. Our attempts to engage with Category 3 scammers proved largely unsuccessful. We collected only a limited number of samples from this category, as these scammers either failed to respond to our engagement attempts or had their phone numbers deactivated.

As explained in Section ~\ref{sec:anatomy}, scammers often redirect conversations to alternative platforms. Of the scammers \sys{} contacted through SMS,  294 conversations were transferred to Telegram and 53 to WhatsApp, as shown in Table~\ref{tab:platform_counts}. While scammers prefer WhatsApp as their handoff platform, we frequently had to negotiate alternative communication methods due to the limitations of the WhatsApp Business API discussed in Section~\ref{subsec:attrition}, resulting in Telegram becoming the more common handoff platform.

\begin{table}[tb]
\centering
\begin{tabular}{lc}
\hline
\textbf{Platforms Used} & \textbf{Number of Conversations} \\
\hline
Text 	$\rightarrow$ Telegram & 294 \\
Telegram Only           & 73 \\
Text $\rightarrow$ WhatsApp          & 53 \\
WhatsApp Only    &  50 \\
Text Only           & 41 \\
Telegram $\rightarrow$ WhatsApp        & 10 \\
\hline
\end{tabular}
\vspace{1ex}
\caption{Platform usage trajectories observed across conversations, showing how scammers transition victims between communication platforms}
\label{tab:platform_counts}
\end{table}

We conduct clustering to identify patterns across scammer messages. To standardize the text for effective clustering analysis, we first preprocess the messages by removing stopwords, person names, and brand names, as these elements are frequently substituted with different values across campaigns while maintaining identical underlying message templates. After this preprocessing, we identify similarities in the initial messages scammers send, including \biggestInitialMessageCluster{} identical messages in the largest cluster. This approach allows us to detect template reuse even when scammers customize messages with different recruiter names or company brands.

Most scammers impersonate well-known companies in their texts to give victims a false sense of security. To identify these companies, we use Named Entity Recognition (NER) through spaCy, a Python library designed for advanced Natural Language Processing (NLP) that provides NER capabilities~\cite{spacyner}. The results of this analysis are shown in Table~\ref{tab:brand-and-names}. The initial messages also mention the alleged recruiter's name, which is actually the scammer's alias. We observe reuse of the same messages and same names, as shown in Table \ref{tab:brand-and-names}.

\begin{figure}[b]
    \centering
    \includegraphics[scale=0.3]{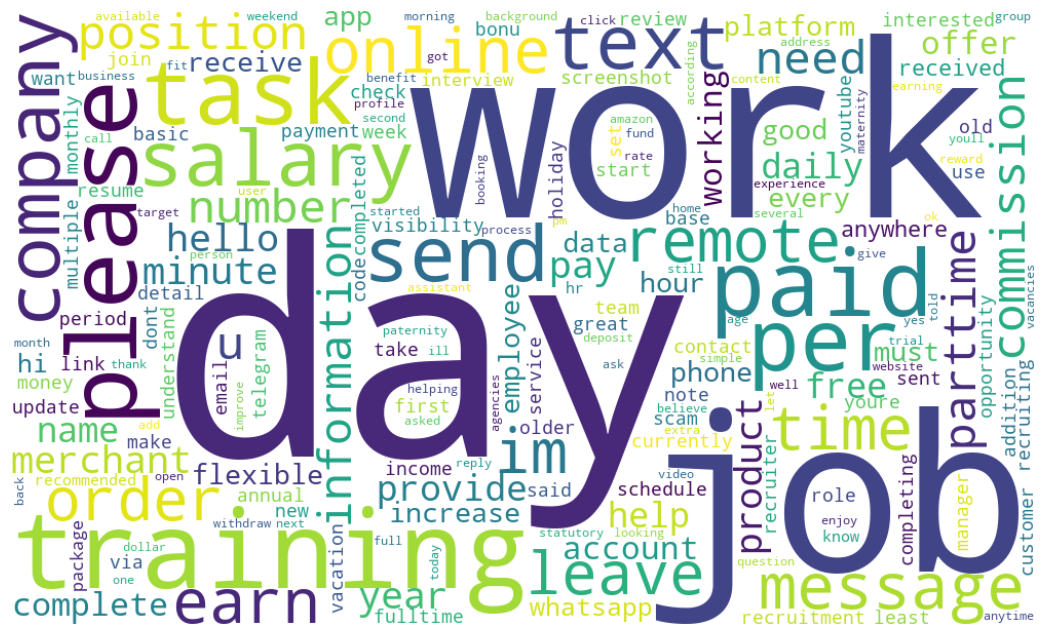} 
    \caption{Most common words sent by scammer}
    \label{fig:wordcloud}
\end{figure}

We analyze patterns and commonalities in messages sent by scammers. The word cloud analysis shown in Figure~\ref{fig:wordcloud} demonstrates that scammers consistently employ professional, employment-related terminology to establish legitimacy. 

We cluster the task details and salary range messages that scammers send and identify identical messages across unique conversations. The most common salary ranges are ``\$50 to \$500" for daily salaries and ``\$6000 to \$15,000", monthly and commissions.  Benefits offered include maternity leave, sick leave, paternity leave, and statutory holidays. 

\begin{figure}[h]
    \centering
    \includegraphics[width=\linewidth]{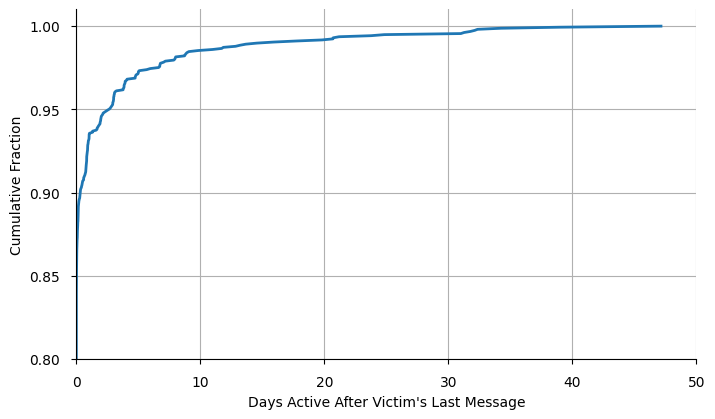} 
    \caption{CDF: Scammer Persistence After Last User Message}
    \label{fig:cdf_scammer_persistence}
    \vspace{1ex}
\end{figure}
\begin{table}
\centering
\caption{Top brand names and personal/platform names mentioned in scam messages detected by \sys}
\label{tab:brand-and-names}
\resizebox{\linewidth}{!}{%
\begin{tabular}{lclc}
\toprule
\textbf{Brand} & \textbf{Count} & \textbf{Name} & \textbf{Count} \\
\midrule
Target                   & \target{}    & Jasmine Martine & \jasmine{}   \\
Amazon                   & \amazon{}    & Judy            & \judy{}      \\
Costco                   & \costco{}    & Maya            & \maya{}      \\
LinkedIn US              & \linkedin{}  & Linda Jackson   & \linda{}     \\
Structube                & \structube{} & Elowen          & \elowen{}    \\
DSL                      & \dsl{}       & Darlene         & \darlene{}   \\
SSENSE Recruiting        & \ssense{}    & Isabella        & \isabella{}  \\
Warner Bros              & \warnerbros{}& Josh Blair      & \joshblair{} \\
Qualtrics                & \qualtrics{} & Monica          & \monica{}    \\
Sony Music Entertainment & \sony{}      & Elizabeth       & \elizabeth{} \\
\bottomrule
\end{tabular}%
}
\end{table}

After obtaining cryptocurrency addresses, we cease communication but continue monitoring incoming messages to analyze scammer persistence. As illustrated in the CDF in Figure ~\ref{fig:cdf_scammer_persistence}, approximately 89\% of scammers stop messaging within just one day of the last user response. By 10 days, around 98\% have ceased contact and by 20 days, approximately 99\% have stopped, with only $\sim 1 \%$ continuing their outreach. A very small group of $\sim 0.5\%$ even attempts contact for over a month. This pattern suggests that while most scammers quickly abandon their efforts when ignored, a persistent minority continues to make outreach over extended periods, likely in an attempt to regain the victim's trust and ultimately defraud them.

\subsection{Domain and IP Analysis}~\label{sec:domain_ip_analysis}

Our analysis of the infrastructure of \uniqueScamWebsites{} unique websites reveals significant clustering. \sys{} resolved these websites to \uniqueIPs{} IP addresses. Upon analysis of hosting providers, we observe that, as shown in Table \ref{tab:ip-vs-providers-sample}, CloudFlare networks host \cloudflarePercent{} of websites, followed by INTEGEN-2 with \integenPercent{}. A significant finding was that 12 websites share the same IP address (192.252.179.27), operated by INTEGEN-2 in Japan. This infrastructure concentration suggests either coordinated operations by the same actors or systematic abuse of specific hosting providers that offer lenient conditions for fraudulent activities. 

\begin{table}[tb]
\centering
\caption{Shared IP addresses hosting multiple scam websites, along with the largest cluster sizes and hosting providers }
\label{tab:ip-vs-providers-sample}
\resizebox{1\linewidth}{!}{%
\begin{tabular}{lrrrr} 
\hline
\textbf{Hosting Provider} & 
\begin{tabular}[c]{@{}r@{}}\textbf{Total Websites} \\  \end{tabular} & 
\begin{tabular}[c]{@{}r@{}}\textbf{Total IPs} \\  \end{tabular} & 
\textbf{Shared IPs} & 
\textbf{Largest Cluster} \\
\hline
CLOUDFLARENET        & 98 & 48 & 17 & 11 \\
INTEGEN-2            & 12 & 1  & 1 & 12 \\
HKUNITED-HK	        & 7  & 1  & 1 & 7 \\
ALIBABA CLOUD - US   & 4  & 3  & 1 & 2 \\
CTG124-25-HK	               & 4  & 1  & 1 & 2 \\
\hline

\end{tabular}
}
\end{table}

\begin{table}[tb]
\centering
\caption{Distribution of top-level domains (TLDs) among scam websites compared to Tranco Top 1M.}
\label{tab:tld-distribution}
\resizebox{0.6\linewidth}{!}{%
\begin{tabular}{lrr} 
\hline
\textbf{TLD} & \textbf{Website \%} & \textbf{Tranco Top 1M \%} \\ 
\hline
.com   & 33.33	\% & 43.96\% \\
.vip   & 16.67\% & 0.31\%  \\
.lat   & 6.86\%  & 0.03\%  \\
.club & 5.88\%  & 0.21\%  \\
.dev   & 3.92\%  & 0.13\%  \\
\hline
\end{tabular}
}
\end{table}

Our analysis of top-level domains (TLDs) used by scam websites reveals a stark contrast with the distribution observed in legitimate web infrastructure. While \textit{.com} remains the most common TLD in both scam and legitimate domains, its proportional use is significantly lower among scams (33.33\%) than in the Tranco Top 1M ~\cite{LePochat2019} list (43.96\%). In contrast, certain TLDs such as \textit{.vip}, \textit{.top}, \textit{.lat}, and \textit{.club} are highly overrepresented in scam infrastructure relative to their appearance in legitimate domains. This aligns with findings from prior industry research, which identifies the aforementioned TLDs as among the most abused for phishing and scam operations due to their low registration costs and lax regulatory oversight~\cite{krebsarticle}~\cite{Sultan_2025}. The percentages of the top 5 TLDs used in our dataset compared to the Tranco Top 1M data are presented in Table~\ref{tab:tld-distribution}.

We analyze scam-related URLs to identify dominant keywords within domain names. The term \textit{ssense} appears most frequently, with 16 occurrences, followed by \textit{structcube} and \textit{coswork}, appearing 9 and 8 times, respectively. Notably, both Ssense and Structube are legitimate retail companies, suggesting that scammers deliberately mimic well-known brands to appear trustworthy. Another recurring name is ``target", referencing the popular retail chain, further illustrating how scammers exploit the credibility of established businesses to deceive victims. 

We discovered that scammers employ domain fronting techniques to evade detection and distribute traffic across multiple fraudulent websites. Our analysis revealed that certain URLs use JavaScript-based redirects to dynamically redirect visitors to different scam websites, a behavior that cannot be detected solely through simple HTTP requests.
To systematically investigate this behavior, we used Selenium to simulate browser interactions and performed multiple visits to suspected fronting domains. A notable example was \texttt{https://target.y***.icu} that initially displays a ``website under maintenance" page but redirects users to different operational scam sites upon reload. Through multiple automated visits to this domain, we observed that each request was redirected to a unique destination.

This dynamic fronting behavior exhibits several key characteristics: All visits are redirected to different subdomains, demonstrating sophisticated traffic distribution mechanisms. The final redirect domains are clustered into two second-level domains: \texttt{y***.cyou} and \texttt{e***.cyou}, which share the same IP, suggesting coordinated infrastructure management. Notably, each redirect consistently routes to the same path (/index/user/login.html), indicating a shared architecture for scam landing pages across the distributed infrastructure.

This technique complicates takedowns by distributing operations across subdomains, evades static URL analysis, and provides redundancy if endpoints are blocked~\cite{nosyk2025exposing}.

\subsection{Clustering}
\subsubsection{IOC Based Clustering} 
~\\
We performed clustering analysis on key infrastructure indicators collected from scammer interactions, including phone numbers, Telegram handles, domains, and cryptocurrency addresses, as presented in Table \ref{tab:scam-indicators}.

\textbf{Domain Clustering:} The most frequently encountered domain was \texttt{wpfm.structube.club}, which appeared in communications from 13 different scammers in our dataset. However, this clustering analysis likely underestimates the true scale of infrastructure sharing as scammers employ domain fronting techniques. As discussed in Section \ref{sec:domain_ip_analysis}, scammers utilize dynamic redirection to distribute traffic across multiple subdomains while maintaining shared backend infrastructure. If domain-fronted domains were considered as single operational units, our clustering results would likely reveal substantially larger connected components.

\textbf{Cryptocurrency Address Clustering:} Our clustering shows how the same Bitcoin and Ethereum addresses are used among different scammers.  However, our investigation revealed that scammers regularly cycle through different cryptocurrency addresses to evade detection and complicate financial tracking. To validate this hypothesis, we conducted a controlled experiment by requesting cryptocurrency addresses from the same scammer at different time intervals. The scammer provided different addresses for each request and explicitly stated: ``Each cryptocurrency transaction address has a 30-minute validity period.'' This finding indicates that cryptocurrency address clustering may not reliably identify connected scam operations, as addresses are deliberately rotated on short timescales.

\begin{table}[tb]
\centering
\caption{Top domains, Telegram usernames, and crypto addresses observed in scam infrastructure}
\label{tab:scam-indicators}
\resizebox{1\linewidth}{!}{%
\begin{tabular}{clc}
\toprule
\textbf{Type} & \textbf{Value} & \textbf{Frequency} \\
\midrule
\multirow{5}{*}{Domain} 
    & wpfm.structube.club & 13 \\
    & tx11.vip & 12 \\
    & target.yy9y.icu & 11 \\
    & ssense.cyou & 7 \\
    & wpfm.structube.lat & 7 \\
\midrule
\multirow{5}{*}{Telegram} 
    & @banshou03 & 4 \\
    & @dabask\_12 & 4 \\
    & @Jensen2008 & 4 \\
    & @Lynn556677 & 3 \\
    & @ssas141242 & 3 \\
\midrule
\multirow{5}{*}{Bitcoin Address} 
    & 36x8XoD8Fu6y5VFk28Qn4tjSSViJ17BsE4 & 7 \\
    & bc1qa6esq4c6wh6ahd6rmla69s32s7wzqkym7m2x77 & 5 \\
    & bc1qk3q0kvlm69nufepd6pn6jz92g6ph8h8g6vfsz6 & 5 \\
    & bc1qqve9cldjcmlx0smsqaqq2nt2nm5vq473n7nc3x & 5 \\
    & 1GBMoUS51c8Qy1BuJyFgTiECkzL8UtkPji & 4 \\
\midrule
\multirow{5}{*}{Ethereum Address} 
    & 0xF3d2554Cc074F52A80DC5115Ce635EBf39b1B26A & 9 \\
    & 0x23B348660d7f54Bc90b672b73C0c3a8c6E9083ff & 8 \\
    & 0x8D37cDdac44a3ad0be4194bF0e74eB773028d376 & 5 \\
    & 0xc041Bf14285d8470b9d1E2464b5042a16566D803 & 4 \\
    & 0xd60170b3123d3a081e883B001b548Ae933B1495b & 4 \\
\bottomrule
\end{tabular}
}
\end{table}

\subsubsection{Web Design Based Clustering}~\label{sec:web_design}
~\\
We identify identical website templates being reused across scam websites. Notably, in our largest cluster, the same homepage template was used across \biggestclusterscreenshot{} websites out of \totalwebsites{}. Furthermore, the invite, notification, and cashback pages also shared templates across 24, 23, and 15 websites in this cluster, respectively.

\begin{figure}[h]
    \centering
     \label{fig:website_cluster_screenshots}
    \includegraphics[scale=0.3]{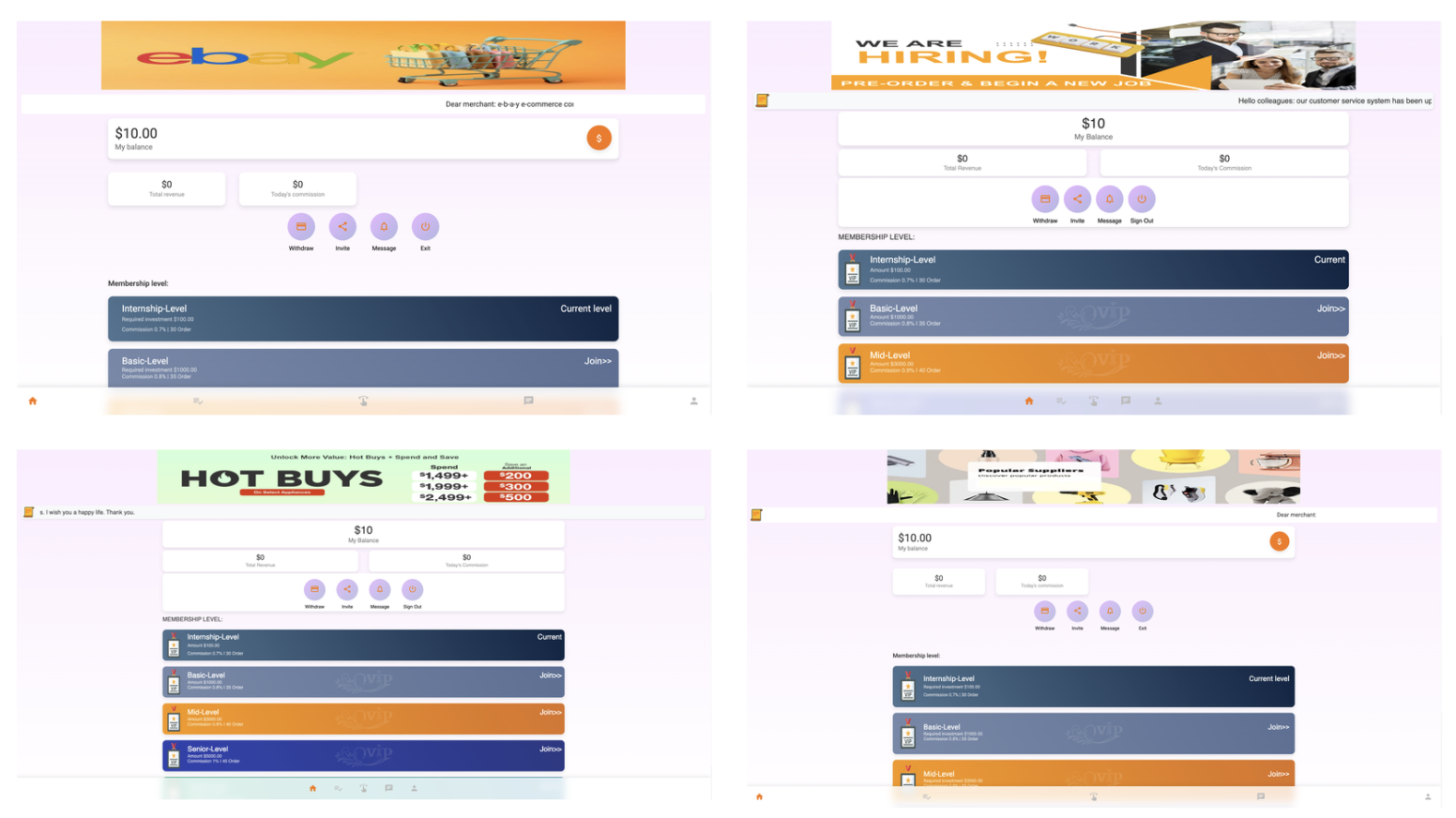} 
    \vspace{1ex}
    \caption{Scam Websites Reusing Identical Web Templates}
    
\end{figure}

\subsubsection{Message Template-Based Clustering}
~\\
We find that scammers reuse the same messaging templates across scams, allowing us to cluster them into several categories. These templates are used for the initial bait message (as seen in Table \ref{tab:top-scam-message-templates}), task descriptions, salary descriptions, and subsequent communication and guidance. Examples of such reused messages are: \textit{``Your training code has been confirmed. Please provide your name and age. This training will be recorded and archived."} which appeared in 113 conversations, \textit{``Now, you need to register a working account on the workbench. New users will get \$10 bonus for registration."} which appeared in 88 conversations and \textit{``We will provide a 60-minute introductory training for everyone to help you better understand the workflow."} which appeared 93 times.

\begin{table}[h]
\centering
\caption{Top Initial Scam Message Templates and Their Occurrence Frequency}
\label{tab:top-scam-message-templates}
\resizebox{\columnwidth}{!}{%
\begin{tabular}{p{7cm}c}
\toprule
\textbf{Initial Message} & \textbf{Frequency} \\
\midrule
Hello! My name is \textit{name} from \textit{brand\_name}. We were really impressed with your profile and would like to provide you the chance to take on a flexible remote role. In this position, you would assist merchants by updating their data... & 600 \\
\midrule
Hello, I'm \textit{name} from the \textit{brand\_name} Recruiting team. Your profile caught our attention through multiple recruiting platforms and we believe you'd be a good fit for our current part-time remote job opportunity. ... & 318 \\
\midrule
Hello, this is \textit{name}, HR Client Service Representative at \textit{brand\_name}. We have obtained your phone number by working with multiple HR platforms. As a result, our company has a job for you as a content promotion assistant, which is a great remote... & 220 \\
\midrule
Hi, I'm \textit{name} from the \textit{brand\_name} Recruiting team. We came across your profile on several recruiting platforms and believe you’d be a great fit for our current remote part-time opportunity. This role involves assisting companies \textit{brand\_name} in.. & 151 \\
\midrule
Hello, my name is \textit{name}, recruiter at \textit{brand\_name}. We came across your profile through several online recruitment platforms and were impressed by your background. We're currently offering a flexible part-time opportunity in your free time... & 103 \\
\bottomrule
\end{tabular} 
}
\vspace{3em}
\end{table}

\subsection{Types of Scams}~\label{sec:types_of_scams}

Our engagement with scammers and scam websites revealed several categories of fraudulent tasks assigned to victims, each designed to appear legitimate. The most common task assigned to victims (\totalTaskScamsPercent{}) involves product review manipulation, where victims are instructed to rate and review products. These tasks create a sense of ``work completion" that justifies the displayed account balance discussed in Section \ref{subsec:wallet_extraction}.

The second most common category involves social media engagement via YouTube (\totalYouTubeScamsPercent{}). Tasks include liking videos, subscribing to channels, and sending screenshots to the scammers. The third most common category (\totalAppStoreScamsPercent{}) involves app store manipulation scams, which operate similarly to the YouTube engagement tasks. In these schemes, victims receive links from scammers that direct them to specific app pages in mobile app stores. Victims are instructed to take screenshots of the app store listings and send the images back to the scammers. The scammers justify these tasks by claiming their company is contracted to increase traffic and visibility for app store listings.

\begin{figure}[h]
    \centering
    
    \includegraphics[scale=1]{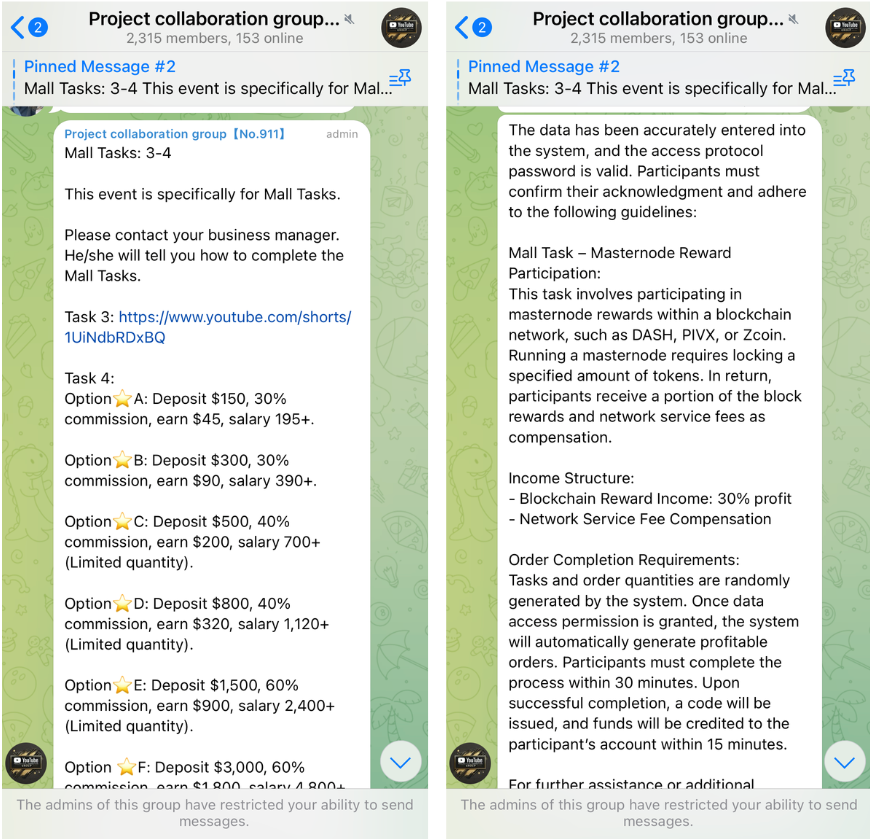} 
    \vspace{1ex}  
    \caption{Scammer Task Description Thread}
    \label{fig:scammerGroup1}
\end{figure}

\begin{figure}[h]
    \centering
    \includegraphics[scale=1]{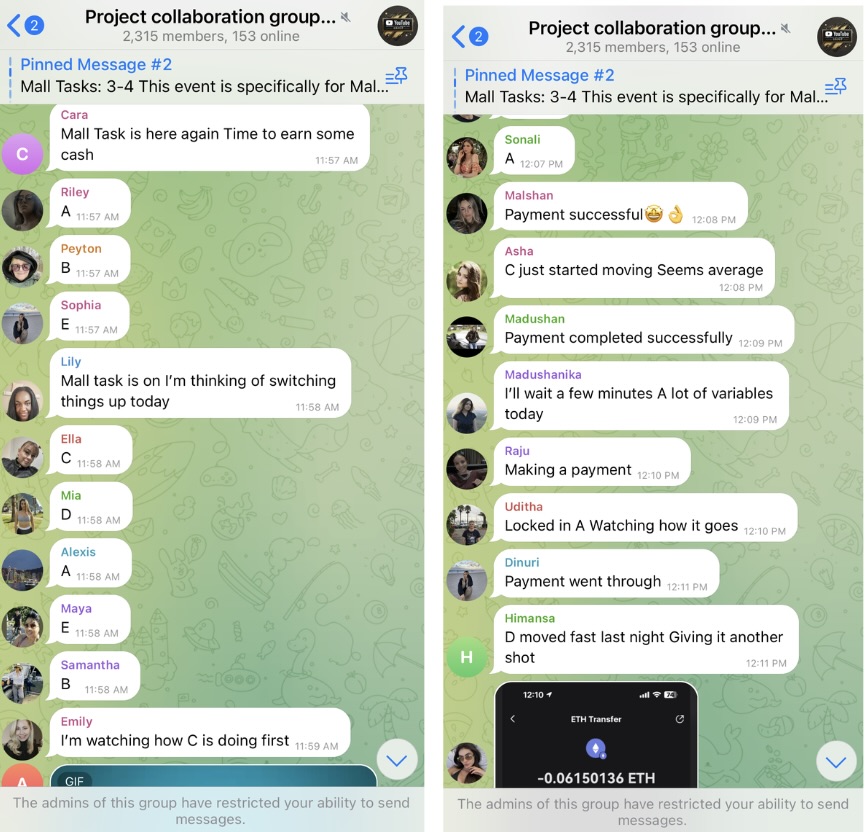} 
    \vspace{1ex}
    \caption{Simulated Employee Participation in Scam Group}
    \label{fig:scammerGroup2}
\end{figure}

Both YouTube and App Store scams are primarily conducted through Telegram, often within groups that appear to contain hundreds of active participants completing tasks and sharing screenshots, as shown in Figure~\ref{fig:scammerGroup1}, which shows the scammer's description of the task, and Figure~\ref{fig:scammerGroup2}, which shows alleged employees engaging in the same tasks. However, we theorize that most of these apparent participants are automated bots designed to create the illusion of a legitimate company with numerous employees, thereby increasing victim confidence in the operation~\cite{sarica_whatsapp_scammers_2023}~\cite{youtube-scam}.

 These scams frequently offer initial payments of \$3 - \$15 for completing tasks, distributed through cryptocurrency platforms such as Coinbase, Cash App, or PayPal's crypto services. Table \ref{tab:crypto-app-counts} provides statistics of the number of scammers mentioning each payment service. The use of regulated platforms like Coinbase~\cite{Coinbase_PrivacyPolicy2024}, which require users to verify their identities with government-issued documents, suggests that scammers either use stolen credentials, forge identity documents to create accounts, or recruit money mules with legitimate accounts. 

Small initial payments build trust, but are merely investments in the larger fraud. Victims eventually encounter ``prepay" tasks that require them to send cryptocurrency first, with scammers claiming this ``optimizes crypto value" and promising returns within hours.  This prepay mechanism represents the primary extraction point where victims lose their funds.

\begin{table}[tb]
\centering
\begin{tabular}{lc}
\hline
\textbf{Payment App} & \textbf{Unique Conversations} \\
\hline
Cash App        & 286 \\
PayPal          & 183 \\
Coinbase        & 54 \\
Crypto.com      & 33 \\
Venmo           & 32 \\
Zelle           & 29 \\
MetaMask        & 15 \\
Strike          & 6 \\
\hline
\end{tabular}
\vspace{1ex}
\caption{Number of Conversations Mentioning Each Payment Platform}
\label{tab:crypto-app-counts}
\end{table}

\subsection{Scammer Profiles}~\label{sec:types_of_scammers}
Our analysis of scammer personas reveals consistent patterns in identity construction and brand impersonation.
Scammers impersonating well-known brands and employ a consistent set of personas. Table~\ref{tab:brand-and-names} presents the most common names and brands used by scammers, with ``Jasmine Martine" being the most frequently mentioned name and ``Target'' the most common brand. Our analysis reveals that most scammers adopt female personas.

We analyzed profile pictures obtained from scammers' Telegram accounts to understand their image sourcing strategies and tested 25 profile pictures using an AI image detector ~\cite{sightengine_detect_ai_images}. Only 3 out of 25 images were flagged as AI-generated. We traced the remaining images to legitimate social media profiles, indicating scammers systematically steal authentic photos to construct believable personas.

%% file: sections/financial-loss.tex
\section{Financial Loss}

\begin{figure}
    \centering
    \includegraphics[scale=0.5]{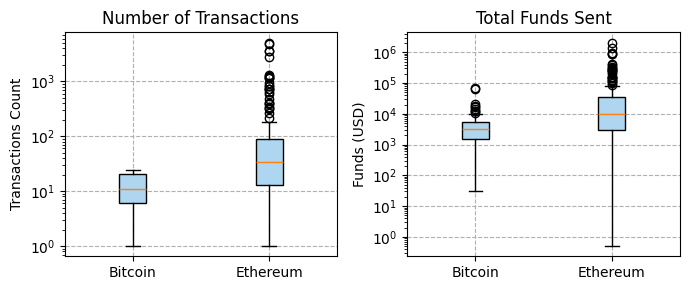}
    \caption{Total number of transactions and US Dollars directed towards scammer-owned cryptocurrency wallets from each victim}\label{fig:numtx_vs_income}
    \vspace{1ex}
\end{figure}

In this section, we discuss the financial losses caused by these scammers and the methodologies we used to calculate them. Since scammers we interacted with primarily used Ethereum and Bitcoin, we focus our analysis on these cryptocurrencies.

During the study period, \sys{} collected \totalBTCwallets{} Bitcoin wallet addresses and \totalETHwallets{} Ethereum wallet addresses from \totalSuccessfulInteractions{} scammer interactions. We observed the reuse of some wallets, resulting in \uniqueBTCwallets{} unique Bitcoin addresses and \uniqueETHwallets{} unique Ethereum addresses.

We employ the approach of Gomez~\etal~\cite{Gomez_2023} to calculate the total amount of bitcoin directed to scammers, using BTC-to-USD and ETH-to-USD historical exchange rates to convert the cryptocurrency amounts into USD on the adjusted closing days when the transactions occurred. We use this approach to prevent overestimation when a portion of the funds is returned to the victim's wallet.

To analyze the financial impact, we traced all incoming transactions to the collected wallet addresses using blockchain explorers and public APIs. For Bitcoin wallets, we used the Blockstream~\cite{blockstream} API, and for Ethereum analysis, we used the Etherscan~\cite{etherscan} API. Transaction data was collected from the wallet creation dates through the end of our study period.

Our methodology involved several key steps to ensure accurate loss calculation. First, we identified all inbound transactions to the scammer-controlled wallets, excluding transactions between wallets within our dataset to avoid double-counting. Second, we filtered out transactions that appeared to be returns or refunds to victims, identified by same-day outbound transactions to different addresses of similar amounts. Third, we applied historical exchange rates from CoinMarketCap.com~\cite{coinmarketcap_bitcoin_historical}~\cite{coinmarketcap_ethereum_historical} to convert cryptocurrency values to USD at the time each transaction occurred.

The analysis revealed significant financial losses for both cryptocurrencies. Bitcoin wallets received a total of \totalBitcoinUSD{} USD, while Ethereum wallets accumulated \totalEthUSD{} USD. The combined financial loss attributed to these scammers totaled \totalLossUSD{} USD during the study period.

We observed notable patterns in transaction behavior across the two cryptocurrencies. Bitcoin transactions showed a median of \bitcoinMedianTxns{} transactions per wallet with an average of \bitcoinAvgTxns{} transactions, ranging from \bitcoinMinTxns{} to \bitcoinMaxTxns{} transactions per address. The median income of Bitcoin per wallet was \bitcoinMedianIncome{}, with an average of \bitcoinAvgIncome{}, ranging from \bitcoinMinIncome{} to \bitcoinMaxIncome{}. In contrast, Ethereum wallets exhibited higher activity levels with a median of \ethMedianTxns{} transactions per wallet and an average of \ethAvgTxns{} transactions, ranging from \ethMinTxns{} to \ethMaxTxns{} transactions. Ethereum income per wallet showed a median of \ethMedianIncome{} with an average of \ethAvgIncome{}, ranging from \ethMinIncome{} to over \ethMaxIncome{} for the most active wallet. We present the top-earning investment scam domains that received payments in Bitcoin and Ethereum, detailing their estimated revenues and operational statuses in Tables~\ref{tab:btc_revenue} and~\ref{tab:eth_revenue}.

\begin{table}[h]
\centering
\caption{Top 20 Scam Domains by Bitcoin Revenue, Number of Transactions and Wallet Details}\label{tab:btc_revenue}
\resizebox{\linewidth}{!}{%
\begin{tabular}{llrr}
\hline
\textbf{Scam Domain} & \textbf{Wallet Address} & \textbf{Total Txns} & \textbf{Revenue} \\ \hline
e3z.ssenen.top, list.ssenusenter.top & bc1qtf...yuqtz & 19 & 69K \\ \hline
riot-blockchains.com & bc1p3u...ex9g... & 11 & 62K \\ \hline
bdog.coswork.click, coswork.app & bc1qf9...usys & 18 & 21K \\ \hline
work.damediasite.com & bc1pj5...hl8a... & 23 & 18K \\ \hline
lktk07o4.1212ll.blog, target.66kk.xyz & bc1qhm...acuk & 16 & 17K \\ \hline
bitdeer-mining.vip, mining-bitdeer.com & bc1p9n...adk0d... & 11 & 15K \\ \hline
u7f1mx.ssbtamika.com & bc1qj8...wqz6 & 21 & 12K \\ \hline
argoblockchain-renew.com & bc1psv...dsfum... & 16 & 12K \\ \hline
tx11.vip & bc1q58...7j03 & 21 & 11K \\ \hline
accio-workbench.com & bc1paq...7ap... & 22 & 10K \\ \hline
tx11.vip & bc1qqv...nc3x & 21 & 9K \\ \hline
dataannotationlaxqw.cc & bc1qh5...qx88 & 19 & 9K \\ \hline
refine-riotplatforms.com & bc1ptg...c79en9q... & 10 & 9K \\ \hline
starsureoak.com & bc1pz9...v4gp... & 14 & 9K \\ \hline
khxylk9k.ssense.dev, ssense.cyou & bc1qct...nf8 & 10 & 8K \\ \hline
webvitality-work.com & bc1qtm...dmsk & 22 & 7K \\ \hline
h5.marm.cloud & bc1p5a...evwhj... & 7 & 7K \\ \hline
ishomedepot.com & bc1qg4...9jwp & 20 & 7K \\ \hline
target.ff9f.xyz, ssense.cyou, ssense.top & bc1qk3...fsz6 & 17 & 6K \\ \hline
new.ebenennav.top & bc1p4z...vejx... & 15 & 6K \\ \hline
\end{tabular}%
}
\end{table}

\begin{table}[h]
\centering
\caption{Top 20 Scam Domains by Ethereum Revenue, Number of Transactions and Wallet Details}\label{tab:eth_revenue}
\resizebox{\linewidth}{!}{%
\begin{tabular}{llrr}
\hline
\textbf{Scam Domain} & \textbf{Wallet Address} & \textbf{Total Txns} & \textbf{Revenue} \\ \hline
pzmsa-751.com, tx11.vip & 0xf3d2...1b26a & 5090 & 2,075K \\ \hline
5c8m8cas.t8tt.xyz, target.ff9f.xyz & 0xc041...6d803 & 4834 & 1,354K \\ \hline
target.bb1b.my, axrt.r9rr.lol & 0xb762...f4ff & 3555 & 933K \\ \hline
lktk07o4.1212ll.blog, target.66kk.xyz & 0x29b9...a0fb & 3531 & 866K \\ \hline
target.x9xx.xyz, j5qz.e8ee.icu & 0x23b3...83ff & 2799 & 846K \\ \hline
nkmqx-793.com & 0xb50f...c163 & 811 & 417K \\ \hline
ratingscore66.com & 0xbc3e...a17a & 1169 & 380K \\ \hline
bitdeer-mining.vip, mining-bitdeer.com & 0xd1f6...e46d & 1266 & 322K \\ \hline
publeicia.com, publeicis.com & 0x3713...b69b & 1185 & 305K \\ \hline
toptai1.com, topta1l.com & 0xdc57...b1a & 718 & 289K \\ \hline
whmkukly.com & 0x64da...8140 & 1297 & 264K \\ \hline
mara-vip.cc & 0x5b40...b993 & 739 & 262K \\ \hline
topta1l.com & 0x8f23...2683 & 477 & 239K \\ \hline
jhyrtui.com & 0x08752...012dc & 1142 & 229K \\ \hline
nclvtrye.com & 0x205b...1593 & 920 & 194K \\ \hline
bdog.coswork.click, coswork.app & 0x4807...302c & 322 & 161K \\ \hline
whmkukly.com & 0xa119...b1c8 & 720 & 161K \\ \hline
bitfarms-boost.com & 0x8594...a00b & 390 & 144K \\ \hline
— & 0xd34a...b626 & 632 & 144K \\ \hline
whmkukly.com, nclvtrye.com & 0x22de...909e & 806 & 130K \\ \hline
\end{tabular}%
}
\end{table}

%% file: sections/discussion.tex
\section{Discussion}~\label{sec:discussion}
\subsection{Coverage}
Due to the ever-growing number of malicious websites, various commercial blocklists exist to warn users about potentially harmful sites. However, due to the large volume introduced daily and to domain fronting techniques discussed previously, many evade detection. Traditional blocklists are optimized to detect malware and phishing websites. Job scams intentionally present legitimate-appearing content that contains no malicious code to trigger detection. This choice is by design: legitimate appearance enables social engineering effectiveness while evading technical detection mechanisms.
To evaluate blocklist effectiveness, we compared our findings against several, as shown in Table \ref{tab:block_lists}. Of the \uniqueScamWebsites{} unique scam websites detected by \sys{}, only 29.6\% were identified by VirusTotal~\cite{virustotal}. This demonstrates that job scams exploit a fundamental gap in security infrastructure, and the diverse nature of scam websites allows many to remain undetected despite existing blocklists.

\begin{table}
\centering
\caption{Percentage of scam websites detected by \sys{} also present in known block-lists.}
\label{tab:block_lists}
\resizebox{0.6\linewidth}{!}{%
\begin{tabular}{lr} 
\hline
\textbf{Name}        & \textbf{Intersection}  \\ 
\hline
VirusTotal~\cite{virustotal}           & 29.6\%                   \\ 
\hline
Metamask~\cite{metamask}             & 7.78\%                    \\ 
\hline
Google Safe Browsing~\cite{GSB} & 15.6\%                    \\ 
\hline
Phishfort~\cite{phishfort_blocklist}            & 0\%                    \\ 
\hline
\end{tabular}
}\
\end{table}

\subsection{Limitations}

\paratitle{Semi-Automation and Human Oversight Requirements} While \sys{} provides significant automation for scammer engagement and data collection, the system currently operates as a semi-automated pipeline that requires human intervention at specific transition points across stages of the pipeline. Human intervention is needed when transitioning from scammer conversations to task completion, where the automated task completion module should first be configured manually with website URLs and referral codes. Similarly, human intervention is required when initiating contact with customer support channels during the wallet extraction phase, and human intervention may be needed during LLM-Assisted Contact. While these steps can be further automated, we refrained from doing so due to the high effort required. Regardless, \sys remains scalable due to natural attrition that occurs as scammers are filtered through each engagement stage. While we initially contact thousands of scammer phone numbers, we narrow it down to a manageable subset that requires human oversight.

\paratitle{Platform Limitations} Currently, Meta does not provide specific provisions for the educational or research use of its API, which presents significant challenges for using WhatsApp as a platform to contact potential scammers for our study. We employed Twilio's WhatsApp Business API as an alternative, but it restricts outreach to unknown recipients ~\cite{twilio2025error63049}. These platform limitations highlight a broader challenge in cybersecurity research: the need for enhanced capabilities that enable researchers to investigate and combat scamming operations more effectively while maintaining appropriate ethical and legal safeguards.

\paratitle{Scam Coverage} In this study, we focus primarily on a specific type of pig-butchering scam: task-related scams. However, similar scams include romance scams, E-ZPass scams, USPS delivery scams, cryptocurrency investment scams, and fake tech support operations. Job scams represent a subset of the broader pig-butchering ecosystem. Additionally, this study was conducted exclusively within the United States, limiting our ability to observe global scam patterns and cross-border operational dynamics. Scam operations often span multiple jurisdictions and may employ diverse tactics, languages, and cultural references when targeting victims in various countries.

%% file: sections/related-work.tex
\section{Related Work}~\label{sec:related-work}

\paratitle{Messaging-Based Fraud Analysis}
Nahapetyan \etal~\cite{nahapetyan2024sms} provide a comprehensive analysis of SMS phishing infrastructure, uncovering systematic abuse in domain registration and hosting ecosystems. 

Agarwal~\etal~\cite{agarwal2024sms} propose a taxonomy of SMS scams based on analysis of millions of messages filtered through the firewalls of mobile operators. They identify eight distinct scam categories and document a 27-fold increase in SMS phishing since 2020.
While their taxonomy provides important categorization frameworks for understanding scam types, their work focuses on classification rather than tracking campaign evolution and scammer behavior patterns across time. Our work complements theirs by providing real-time behavioral clustering and longitudinal analysis of how individual campaigns evolve.

Additional work has explored broader online fraud ecosystems. Kotzias~\etal~\cite{ctrl_alt_deceive} quantify user exposure to online scams across multiple platforms, while Kotzias~\etal~\cite{scamdog_millionaire} focus on e-commerce fraud detection. However, these studies do not address the specific communication patterns and multi-stage engagement tactics characteristic of pig butchering scams.

\paratitle{Cryptocurrency Scam Detection Systems}
CryptoScamTracker~\cite{double_or_nothing} focuses on identifying cryptocurrency giveaway scams that promise to ``double" any cryptocurrency sent to fraudulent addresses. These scams often exploit social media platforms and utilize celebrity impersonation to establish credibility.

Muzammil~\etal~\cite{muzammil2025babylon} introduce \textit{Crimson}, a real-time detection system for cryptocurrency investment scam websites. Their longitudinal analysis of over 43,000 scam domains reveals clustering patterns based on web design, hosting infrastructure, and reused identifiers such as wallet addresses and social media handles. While their focus on web-based infrastructure differs from our SMS-centric approach, their methodology for large-scale behavioral clustering and infrastructure mapping provides valuable insights.

Additional work has explored AI-assisted detection of phishing emails~\cite{phishing-emails}  and comment bot detection on platforms like YouTube \cite{evolving_bots}.  Roy\~etal~\cite{chatbots_to_phishbots} examine how LLMs can be exploited for generating phishing scams, highlighting emerging threats in AI-enabled fraud.

\paratitle{Direct Scammer Interaction Studies} Miramirkhani~\etal~\cite{Miramirkhani_2017} conducted telephone-based interactions with scammers to study social engineering techniques, while Li~\etal~\cite{youtube-scam} engaged scammers on YouTube to reveal operational strategies. While these studies provide crucial insights unavailable through passive monitoring, they face significant scalability limitations due to manual engagement.
While these direct engagement studies provide crucial insights that cannot be obtained through passive monitoring, they are constrained by scalability limitations arising from their manual nature. Our approach addresses this limitation by automating the process of engaging scammers using large language models, enabling systematic interaction with scammers at scale while preserving the depth of insight provided by direct engagement.

%% file: sections/conclusion.tex
\section{Conclusion}~\label{sec:conclusion}
In this paper, we developed and deployed \sys{} to systematically engage with and analyze job-based smishing scams over a 10-month period. Our system collected \totalMessages{} messages from \totalScammersNumbers{} scammer phone numbers across SMS, WhatsApp, and Telegram platforms, successfully engaging with over 1900 scammers through automated LLM-driven conversations. We conducted a comprehensive analysis of scammer operations, clustering them based on message templates, infrastructure reuse, cryptocurrency wallets, and behavioral patterns. Our findings revealed extensive coordination among scam operations, with identical message templates appearing across hundreds of conversations and significant clustering of infrastructure among specific hosting providers. We identified three primary categories of fraudulent tasks and documented the progression of social engineering from small, legitimate payments to large cryptocurrency extraction schemes. In terms of financial losses, we estimated \totalLossUSD{} sent by victims to scammer-controlled Bitcoin and Ethereum wallets, with individual operations ranging from hundreds of dollars to over \$2 million.

%% file: sections/ethics.tex
This research was conducted under the oversight and approval of our institution's Institutional Review Board (IRB), which reviewed and approved all aspects of our methodology involving human subject interactions. Given that our study required active engagement with individuals engaged in fraudulent activities, we obtained comprehensive IRB approval to ensure that our research methods met established ethical standards for human subjects research.
Our experimental design poses minimal risk to participants, as we exclusively simulate authentic victim behavior that scammers would naturally encounter in their typical operations. By adopting realistic victim personas and following standard interaction patterns, our approach mirrors the experiences that these individuals would have with genuine targets, without introducing any novel risks or harms beyond their existing operational exposure.
The research methodology necessitated that participants (scammers) remain unaware of the true research purpose to preserve the authenticity and validity of their responses. Disclosure of our research objectives would fundamentally compromise data integrity by altering participant behavior and potentially rendering findings non-representative of actual scam operations. To address this methodological requirement, we obtained an informed consent waiver from the IRB, which determined that the research met established criteria for waiver approval, including minimal risk to participants and the impracticality of conducting meaningful research with prior disclosure.

%% file: sections/appendix.tex
\begin{figure}[h]
    \centering
    \includegraphics[scale=0.5]{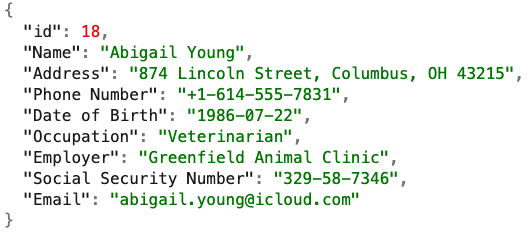} 
    \caption{Example of a synthetic victim persona generated by \sys{} to ensure authenticity and consistency when LLM agents engage with scammers across shifting platforms} 
    \label{fig:llm_persona}
\end{figure}